\documentclass[aps,pra,11pt]{revtex4-2}
\usepackage[english]{babel}
\usepackage{amsmath,amsthm,amssymb,graphicx,mathtools}
\usepackage[colorlinks=true, allcolors=blue]{hyperref}
\usepackage{xcolor}
\usepackage{physics}
\usepackage{hyperref,braket}
\usepackage{physics}

\newcommand{\mga}[1]{\textcolor{black}{#1}}
\newcommand{\jy}[1]{\textcolor{black}{#1}}

\begin{document}
\title{\mga{Fisher geometry reshapes the effect of incompatibility in multiparameter quantum estimation}}
\author{Jiayu He}
\affiliation{Department of Physics, University of Helsinki, FI-00014 Helsinki, Finland}
\author{Matteo G. A. Paris}
\affiliation{Dipartimento di Fisica,  Universit\`{a} di Milano, I-20133 Milano, Italy}

\begin{abstract}
\mga{Multiparameter quantum estimation faces two fundamental obstacles: sloppiness, i.e., anisotropy of the quantum Fisher information matrix (QFIM) that renders some parameter directions insensitive, and incompatibility, the non-commutativity of optimal measurements for different parameters. The trade-off bound $C_T$ captures their joint impact on precision, but it has remained unclear how the distribution of incompatibility across parameter planes affects its overall cost. Here we separate the total amount of incompatibility from its location. We introduce a dimensionless quantity $G_n^{(F)}$ that measures the alignment between the incompatibility distribution and the eigenvalues of the QFIM, and show how the Frobenius scale of the incompatibility contribution factorizes. We obtain a bound and prove the incompatibility cost lies between this bound and a rank-dependent multiple thereof.  We also prove that at fixed \jy{sloppiness, or equivalently fixed Fisher volume}, concentrating incompatibility into a single parameter plane reduces the optimized trade-off cost because the Fisher geometry can then be reshaped to allocate more Fisher area to that plane. A qutrit $SU(2)$ encoding numerically confirms that states with larger incompatibility strength can nevertheless incur a smaller cost if the matching factor $G$ is sufficiently small. Our results establish that the distribution of incompatibility relative to the Fisher eigenbasis is a central diagnostic for multiparameter estimation, beyond the total incompatibility strength.}
\end{abstract}
\maketitle

\section{Introduction}
\mga{
Quantum estimation theory has long provided a powerful framework for achieving fundamental precision limits \cite{helstrom1967, holevo2011, braunstein1994,Paris2009,Caves1,Caves2,giovannetti2004,Giovannetti2006}. In single-parameter scenarios, the quantum Cramér-Rao bound is attainable, and the optimal measurement is often well understood \cite{helstrom1967,holevo2011,braunstein1994, Paris2009}. However, realistic quantum sensors, clocks, and imaging systems almost invariably involve multiple unknown parameters—magnetic field components, phases, frequencies, or geometric angles \cite{Szczykulska2016, Liu2020, Demkowicz2020,albarelli2020perspective}. In such multiparameter settings, two distinct effects degrade precision. First, the Fisher information matrix may be highly anisotropic (“sloppiness”) \cite{he2025,Frigerio2025, Yang2025}: some parameter combinations are far more sensitive than others, limiting the effective information available. Second, the optimal measurements for different parameters may be incompatible, preventing simultaneous saturation of the quantum Cramér-Rao bound \cite{Nagaoka1989,Suzuki2016,Yamagata2013,Yuen1973, ragy2016Compatibility, Matsumoto2002}. Understanding how these two effects interact is not just a theoretical question—it is central to designing practical multiparameter estimation schemes, from quantum magnetometry to phase–superresolution imaging~\cite{Hou2020,Hou2021,Humphreys2013,Pezze2017,Tsang2016}.}

\mga{The trade-off bound $C_T$ has recently emerged as a relevant figure of merit that quantifies how sloppiness and incompatibility jointly constrain precision \cite{he2025weight,albarelli2020perspective}. Yet a key gap remains. The bound depends on the matrix $M=Q^{-1}\,U\,Q^{-1}$, where $Q$ is the quantum Fisher information matrix (QFIM) and $U$ the Uhlmann curvature. It has been natural to ask: does the cost depend only on the total amount of incompatibility, or does the distribution of incompatibility across different parameter planes matter? In this work, we show decisively that distribution matters. Two estimation problems with identical total incompatibility can have very different values of $C_T$ if the incompatibility is concentrated in parameter planes with large or small Fisher area. This insight turns a crude {\em how much} question into a refined {\em where} one, with direct implications for probe-state design and measurement strategies.}

\mga{
Our analysis reveals a subtle interplay: the Fisher geometry itself can be reshaped—subject to a fixed determinant constraint—to mitigate the cost of incompatibility. When incompatibility is concentrated in a single parameter plane, the optimal strategy reduces the corresponding Fisher eigenvalue, thereby increasing the Fisher area of that plane and suppressing the cost term. In other words, sloppiness can be reorganized to accommodate incompatibility. This is not obvious a priori: one might have expected that concentrating incompatibility worsens the bound. Instead, we find that concentration enables targeted geometric countermeasures. A three-parameter qutrit example confirms that states with larger normalized incompatibility can nevertheless yield a smaller total cost if the matching factor is small, i.e., \jy{if the Fisher geometry allocates a large Fisher area to the dominant incompatible one}. These results provide a new diagnostic for probe-state optimization and open the door to designing multiparameter sensors that are not only {\em compatible enough} but also geometrically well-matched.}

\section{Preliminaries}
\subsection{Quantum Geometric Tensor and Multiparameter Estimation}
A quantum statistical model can be viewed as a map from a parameter manifold to the space of quantum states: $\boldsymbol\lambda\longmapsto \rho_{\boldsymbol\lambda}$. For a pure-state model this becomes $\boldsymbol\lambda\longmapsto |\psi_{\boldsymbol\lambda}\rangle$. Thus each tangent direction \(\partial_\mu\) in parameter space is mapped
to a tangent vector \(|\partial_\mu\psi\rangle\) in Hilbert space. Since physical pure states are rays, the relevant geometry is the geometry of projective Hilbert space. For pure-state models, the quantum geometric tensor is the pullback of this geometry to the parameter manifold \cite{provost1980riemannian}:
$$\mathcal Q_{\mu\nu}=\langle \partial_\mu\psi|(1-|\psi\rangle\langle\psi|)|\partial_\nu\psi\rangle$$.
Its real and imaginary parts have different meanings: $ \mathcal Q_{\mu\nu} = g_{\mu\nu} + i\Omega_{\mu\nu}$. The real part $g_{\mu\nu} =\operatorname{Re}\mathcal Q_{\mu\nu}$ defines a Riemannian metric on parameter space. In quantum estimation, this metric is proportional to the QFIM, $Q_{\mu\nu}=4g_{\mu\nu}$. Thus the infinitesimal squared distance induced on parameter space is
\begin{equation*}
    ds^2=\sum_{\mu,\nu} g_{\mu\nu}\,d\lambda_\mu d\lambda_\nu=\frac14\sum_{\mu,\nu}
    Q_{\mu\nu}\,d\lambda_\mu d\lambda_\nu .
\end{equation*}
Therefore, the QFIM measures the sensitivity of the quantum state to small parameter displacements. Large eigenvalues of $Q$ correspond to directions in parameter space along which the quantum state changes rapidly, while small eigenvalues correspond to poorly encoded parameter directions.

The imaginary part of the QGT is the Berry curvature $\Omega_{\mu\nu}=\operatorname{Im}\mathcal Q_{\mu\nu}$. With the normalization used below, we define
\begin{equation}
    U_{\mu\nu}=4\Omega_{\mu\nu}=4\operatorname{Im}\mathcal Q_{\mu\nu}.
    \label{eq:U_from_QGT}
\end{equation}
Geometrically, $U_{\mu\nu}$ measures the curvature on the parameter plane $(\mu,\nu)$. It describes the geometric phase, or the nontrivial geometric structure, generated by moving the quantum state along two different parameter directions.

The same QFIM also has an operational definition in local quantum estimation theory. For a general mixed-state model $\boldsymbol\lambda \longmapsto \rho_{\boldsymbol\lambda}$, the symmetric logarithmic derivative (SLD) $L_\mu$ associated with the
parameter \(\lambda_\mu\) is defined by \cite{helstrom1967}
\begin{equation}
    \partial_\mu \rho_{\boldsymbol\lambda}=\frac12
    \left(\rho_{\boldsymbol\lambda} L_\mu+L_\mu \rho_{\boldsymbol\lambda}\right).
    \label{eq:SLD_definition}
\end{equation}
The corresponding QFIM is
\begin{equation}
    Q_{\mu\nu}=\frac12 \operatorname{Tr}\left[ \rho_{\boldsymbol\lambda} \{L_\mu,L_\nu\} \right].
    \label{eq:SLD_QFIM}
\end{equation}
This definition gives $Q$ its operational meaning: it determines the compatible precision scale through the SLD quantum Cramér--Rao bound,
\begin{equation}
    \operatorname{Cov}(\hat{\boldsymbol\lambda})\geq Q^{-1},
    \label{eq:SLD_QCRB}
\end{equation}
whenever $Q$ is invertible.

In the SLD formulation, the incompatibility between different parameter directions is captured by the commutators of the SLDs. With the convention used in this work, we write the Uhlmann curvature matrix
\begin{equation}
    U_{\mu\nu}=\frac{1}{2i}\operatorname{Tr} \left[\rho_{\boldsymbol\lambda}[L_\mu,L_\nu]\right].
    \label{eq:SLD_incompatibility_matrix}
\end{equation}
The weak compatibility condition is \cite{ragy2016Compatibility}
\begin{equation}
    \operatorname{Tr}
    \left[\rho_{\boldsymbol\lambda} [L_\mu,L_\nu] \right] = 0
    \qquad
    \text{for all } \mu,\nu .
    \label{eq:weak_compatibility_condition}
\end{equation}
When this condition is violated, the SLD quantum Cramér--Rao bound cannot, in general, be saturated simultaneously for all parameters.

For pure-state models, the SLD formulation and the QGT formulation are consistent. The SLD-QFIM coincides with the metric part of the pure state QGT, while the Uhlmann curvature matrix corresponds to the curvature part, up to the chosen normalization convention. The QGT provides the geometric origin of the metric and curvature, while the SLD formulation gives their operational meaning in local estimation theory.

In the following, we use the rescaled quantum geometric tensor
\begin{equation}
    F_{\mu\nu}:=4\mathcal Q_{\mu\nu}=Q_{\mu\nu} + iU_{\mu\nu}.
    \label{eq:rescaled_QGT}
\end{equation}
Equivalently, in matrix form, $F = Q+iU$, where $Q^T=Q\geq0$, $U^T=-U$. Since $F$ is a Gram matrix of projected tangent vectors, it must be positive semidefinite:
\begin{equation}
    F=Q+iU\geq0.
    \label{eq:QGT_feasibility}
\end{equation}
This condition represents the QGT {\em feasibility condition}, showing that $Q$ and $U$ cannot be chosen independently.

\subsection{Compatible and incompatibility-corrected precision bounds}
\label{subsec:compatible_incompatibility_corrected_bounds}
In multiparameter estimation, the covariance matrix of an estimator is a matrix-valued quantity. To obtain a scalar precision cost, we introduce a positive weight matrix $W$, which specifies the relative importance of different parameters or parameter combinations. In the following, we omit the overall factor $1/M$, where $M$ is the number of independent repetitions.

The scalar form of the SLD quantum Cramer--Rao bound is $C_S[W]=\operatorname{Tr}[WQ^{-1}]$. This bound depends only on the metric part $Q$. It therefore provides the compatible precision baseline: it is the cost one would obtain if the SLD bound were simultaneously attainable for all parameters.

In general multiparameter estimation, however, the SLD bound need not be attainable because the locally optimal measurements for different parameters may be incompatible. The asymptotically attainable bound is the Holevo bound \cite{holevo2011}
\begin{equation*}
    C_H[W]=\min_{X\in\mathcal X}\left\{\operatorname{Tr}
    \left[ W\,\operatorname{Re} Z[X]\right]+\left\|
    \sqrt W\,\operatorname{Im} Z[X]\,\sqrt W\right\|_1\right\},
\end{equation*}
where $Z_{\mu\nu}[X]=\operatorname{Tr}[\rho_{\boldsymbol\lambda}X_\mu X_\nu]$, and the minimization is over all operators \(X_\mu\) satisfying the local unbiasedness constraints. Although $C_H[W]$ is the fundamental multiparameter precision bound, it usually involves a nontrivial operator optimization and is difficult to evaluate analytically. A useful explicit upper bound is obtained by choosing $\widetilde X_\mu=\sum_\nu (Q^{-1})_{\mu\nu}L_\nu$ in the Holevo optimization \cite{he2025weight, albarelli2020perspective}:
\begin{equation}
 C_T[W]:= C_S[W]+\left\|\sqrt W Q^{-1}UQ^{-1}\sqrt W \right\|_1 .
    \label{eq:CT_bound}
\end{equation}
Thus $C_T$ is an explicit incompatibility-corrected upper bound to the Holevo bound, $C_H[W]\leq C_T[W]$. The first term $C_S[W]=\operatorname{Tr}[WQ^{-1}]$ is the compatible metric cost. The second term is the incompatibility term,
\begin{equation}
    C_T[W]-C_S[W] = \left\|\sqrt W Q^{-1}UQ^{-1}\sqrt W\right\|_1 .
    \label{eq:incompatibility_cost_weighted}
\end{equation}
It is convenient to introduce the relative incompatibility term
$$T[W] := \frac{\left\|\sqrt W Q^{-1}UQ^{-1}\sqrt W  \right\|_1 }{\operatorname{Tr}[WQ^{-1}] }.$$
Then $C_T[W]=(1+T[W])C_S[W]$, so $T[W]$ measures how large the incompatibility term is relative to the compatible SLD cost. 

A coarser but weight-independent way to bound the same contribution is
obtained from
$$
    \left\| \sqrt W Q^{-1}UQ^{-1}\sqrt W  \right\|_1\leq \operatorname{Tr}[WQ^{-1}] \, \left\|iUQ^{-1}\right\|_\infty .
$$
This motivates the quantumness measure \cite{quantumness2019, razavian2020quantumness}
\begin{equation}
  R := \left\|iUQ^{-1}\right\|_\infty .  
  \label{R}
\end{equation}
    
Therefore $T[W]\leq R$, and the bounds satisfy the hierarchy
\begin{align*}
		C_{S}[W]\leq C_H[W]\leq (1+T[W])C_{S}[W]\leq (1+R)C_{S}[W]\,.
\end{align*}

The quantity $R$ gives a broad, weight-independent upper bound on the relative gap, while $T[W]$ keeps the dependence on the chosen precision task through $W$. In the rest of this work, we focus on the structure of the incompatibility part. In particular, for $W=\mathbf 1$, the central object becomes
\begin{equation}
    C_T-C_S = \left\|Q^{-1}UQ^{-1}\right\|_1 .
    \label{eq:CT_minus_CS_W_identity}
\end{equation}
This expression shows explicitly how the Uhlmann curvature matrix $U$ is converted into a precision cost by the inverse QFIM $Q^{-1}$.

\subsection{Measures of sloppiness and incompatibility}

We use two scalar quantities to characterize the model. The first is the sloppiness measure \cite{he2025}
\begin{equation}
    s:=\frac{1}{\det Q}= \det(Q^{-1}),
 \label{eq:sloppiness_measure}
\end{equation}
which is the inverse Fisher volume. A large value of $s$ indicates a small overall Fisher volume and therefore poor distinguishability in the parameter space.

The second is the incompatibility measure \cite{he2025}
\begin{equation}
    c:=\frac12\operatorname{Tr}\left[ U^\dagger U\right]=\frac12\|U\|_F^2,
    \label{eq:incompatibility_measure}
\end{equation}
where $\|\cdot\|_F$ is Frobenius norm. Since $U$ is antisymmetric, this can be written as $c =\sum_{\mu<\nu}U_{\mu\nu}^2.$ Thus $c$ measures the total pairwise incompatibility among the parameters and vanishes when the weak compatibility condition is satisfied. 

In the following, we use $s$ and $c$ as compact diagnostics of two different sources of precision loss: $s$ characterizes the loss caused by overall distinguishability scale, while $c$ characterizes the total amount of incompatibility, which originates from the noncommutativity of the locally optimal quantum measurements.

\section{A geometric bound on the incompatibility}

This section focuses on the incompatibility contribution appearing in the explicit bound $C_T$. We first recall the weight-independent quantumness $R$ and the weight-dependent quantity $T[W]$, where $R$ characterizes a model-level incompatibility scale, while \(T[W]\) resolves the incompatibility contribution for a chosen precision cost. In what follows,  we focus on the incompatibility contribution $C_T-C_S=\|Q^{-1}UQ^{-1}\|_1.$ The goal is to understand the structure of this term. The main point is that the total amount of incompatibility is not sufficient to determine its contribution to $C_T$. The same amount of incompatibility can lead to different costs depending on how its pairwise components are placed relative to the Fisher geometry. 

\subsection{$R$, $T$, and Fisher geometry of incompatibility}

The quantumness $R$ provides a  measure of multiparameter incompatibility. Its geometric origin can be seen from the QGT feasibility condition. Define the Fisher-normalized incompatibility matrix
$$B:=Q^{-1/2}UQ^{-1/2}.$$
Then $Q+iU=Q^{1/2}(I+iB)Q^{1/2}$. Since $Q^{1/2}$ is invertible, the QGT positivity condition $Q+iU\geq0$ is equivalent to
$$I+iB\geq0.$$

Because $B$ is real skew-symmetric, its eigenvalues are of the form $\{\pm i\beta_1,\ldots,\pm i\beta_r\}$, with $r=\lfloor n/2\rfloor$ and $\beta_\alpha\geq0$. Equivalently, the $\beta_\alpha$'s are the singular values of $B$. The eigenvalues of canonical two-dimensional blocks $I+iB_\alpha$ are $1\pm\beta_\alpha$. Therefore based on $ I+iB\geq0$, we have 
\begin{equation}
    \sigma_{\max}\left(B\right)\leq1.
    \label{eq:QGT_feasibility_B}
\end{equation}

This is the geometric content of the quantumness $R$. Since $iUQ^{-1}$ is similar to $iQ^{-1/2}UQ^{-1/2}=iB$, $R$ measures the largest Fisher-normalized incompatibility mode. Thus, with the usual spectral interpretation,
$$R= \|iUQ^{-1}\|_\infty = \|iB\|_\infty=\sigma_{\max}(B).$$

To study the structure of $T[W]$, it is useful to isolate the matrix that enters the incompatibility contribution. For general $W$, define
$$M_W:=\sqrt W Q^{-1}UQ^{-1}\sqrt W .$$
Then $C_T[W]-C_S[W]=\|M_W\|_1$. The quantity $R$ is naturally related to the Fisher-normalized matrix $Q^{-1/2}UQ^{-1/2}$, which also appears in the QGT feasibility condition. By contrast, the $T$-term contains $Q^{-1}UQ^{-1}$, where each pairwise incompatibility is divided by the full Fisher area of the corresponding parameter plane.

In the rest of this work, we take $W=\mathbf 1$ in order to expose the geometric mechanism without the additional distortion introduced by the weight matrix. We then write
$$ M:=Q^{-1}UQ^{-1}, \qquad C_T-C_S=\|M\|_1.$$

The difference between $B$ and $M$ is transparent in the eigenbasis of $Q$. If $Q=\operatorname{diag}(\lambda_1,\ldots,\lambda_n)$, then
$$
    B_{ij} =\frac{U_{ij}}{\sqrt{\lambda_i\lambda_j}},  \qquad
    M_{ij} =\frac{U_{ij}}{\lambda_i\lambda_j}.
$$
Thus $B$ compares the pairwise incompatibility $U_{ij}$ with the Fisher-area amplitude $\sqrt{\lambda_i\lambda_j}$. This is why $B$ naturally appears in the QGT feasibility condition and in the quantumness $R$. By contrast, $M$ compares $U_{ij}$ with the full Fisher area $\lambda_i\lambda_j$ of the parameter plane $(i,j)$. Therefore $M$ measures the incompatibility per Fisher area, which is the quantity that enters the precision penalty $C_T-C_S$.

\subsection{Fisher-area bound on the incompatibility}

To obtain a more transparent geometric scale, we compare $\|M\|_1$ with the Frobenius norm:
$$
    \|M\|_F\leq\|M\|_1\leq\sqrt r\,\|M\|_F ,
$$
where $r=\operatorname{rank}(M)$ and $r$ is even. Thus the Frobenius norm gives a lower bound and, up to the rank factor, an upper bound on the exact incompatibility term.

We now evaluate $\|M\|_F$ in the eigenbasis of $Q$. Since $M$ is skew-symmetric, its Frobenius norm is
\begin{equation}
    \|M\|_F^2=2\sum_{i<j}\frac{U_{ij}^2}{\lambda_i^2\lambda_j^2}.
    \label{eq:MF_pairwise_s_form}
\end{equation}
This expression shows explicitly that each pairwise incompatibility $U_{ij}$ is divided by the Fisher area $\lambda_i\lambda_j$ of the same parameter plane. 

To separate the total amount of incompatibility from its placement among different parameter planes, define
$$
    \tilde U_{ij}:=\frac{U_{ij}^2}{c}, \qquad \sum_{i<j} \tilde U_{ij}=1,
$$
where $c=\frac12\|U\|_F^2=\sum_{i<j}U_{ij}^2$ measures the total pairwise incompatibility, while $\tilde U_{ij}$
describes how this incompatibility is distributed among the parameter planes.

We also remove the overall Fisher-information scale using the sloppiness $s=\frac{1}{\det Q}$. The dimensionless Fisher eigenvalues are
$$
    \tilde\lambda_i:=\frac{\lambda_i}{\sqrt[n]{\det Q}}=s^{1/n}\lambda_i, \qquad\prod_{i=1}^n\tilde\lambda_i=1.
$$
The dimensionless Fisher area of the parameter plane $(i,j)$ is $\tilde\lambda_i\tilde\lambda_j$. This motivates the Fisher-area matching factor
\begin{equation}
    G_n^{(F)}:=\sum_{i<j} \tilde U_{ij}\frac{1}{\tilde\lambda_i^2\tilde\lambda_j^2}.
    \label{eq:GnF_def_s_form}
\end{equation}
This quantity is large when the dominant incompatibility components lie in parameter planes with small relative Fisher area, and small when they are supported by large relative Fisher areas.

Substituting $U_{ij}^2=c\,\tilde U_{ij}$ and $\lambda_i\lambda_j=s^{-2/n}\tilde\lambda_i\tilde\lambda_j$ into
Eq.~\eqref{eq:MF_pairwise_s_form}, we obtain
\begin{equation}
    \sqrt{2c}\,s^{2/n}\sqrt{G_n^{(F)}} \leq C_T-C_S \leq \sqrt r\, \sqrt{2c}\,s^{2/n}\sqrt{G_n^{(F)}} .
    \label{eq:CTCS_bound_s_G}
\end{equation}
Eq~\eqref{eq:CTCS_bound_s_G} shows that the exact trace-norm contribution $C_T-C_S$ is trapped between the Frobenius scale and its rank-dependent multiple. Hence $\sqrt{2c}\,s^{2/n}\sqrt{G_n^{(F)}}$ sets the size of incompatibility contribution up to the factor $\sqrt{r}$. In this sense, decreasing this scale simultaneously lowers both the lower and upper bounds on $\|M\|_1$.
The analytical upper bound of incompatibility contribution is controlled by three factors: the total incompatibility strength $\sqrt c$, the sloppiness scale $s^{2/n}$, and the Fisher-area matching factor $\sqrt{G_n^{(F)}}$. The role of $G_n^{(F)}$ is to describe how the incompatibility is placed relative to the Fisher geometry. A large $G_n^{(F)}$ means that the dominant incompatibility components lie in parameter planes with small relative Fisher area. A small $G_n^{(F)}$ means that the dominant incompatibility is supported by parameter planes with large Fisher area. Therefore, the total amount of incompatibility $c$ alone does not
determine the precision penalty.

For three parameters, the Frobenius scale becomes exact. A $3\times3$ skew-symmetric matrix has only one nonzero singular-value pair, so $\|M\|_1=\sqrt2\,\|M\|_F$. Writing
$\mathbf u=(U_{23},U_{31},U_{12})^T$, one has $c=|\mathbf u|^2$. In this case,
$$
    G_3^{(F)} =\sum_{k=1}^3\tilde u_k^2\tilde\lambda_k^2, \qquad
    \tilde u_k^2=\frac{u_k^2}{|\mathbf u|^2}.
$$
Therefore,
\begin{equation}
    C_T-C_S = 2\sqrt c\,s^{2/3}\sqrt{G_3^{(F)}} .
    \label{eq:CTCS_three_parameter_s_form}
\end{equation}

\section{Quantum Fisher information allocation at fixed sloppiness}
\label{sec:fisher_allocation_fixed_sloppiness}
We now turn from the cost of a fixed Fisher geometry to the question of how Fisher information should be allocated at fixed sloppiness. The compatible case $U=0$ provides a reference point: the optimal Fisher eigenvalues are determined only by the weight matrix $W$ and the minimum cost scales with the sloppiness $s=1/\det Q$. When $U\neq0$, the incompatibility term changes this picture. The effect is simple in two parameters, where there is only one incompatible parameter plane, but becomes genuinely distribution-dependent in three parameters.

\subsection{The compatible case at fixed sloppiness}

We first consider the compatible case. Assume that $Q$ and $W$ are simultaneously diagonal in the working basis, 
$Q=\operatorname{diag}(\lambda_1,\ldots,\lambda_n)$, $W=\operatorname{diag}(w_1,\ldots,w_n)$, with $w_k>0$. Then $C_S[W]=\sum_k w_k/\lambda_k$. We minimize this
quantity at fixed sloppiness,
$$
    s=\frac{1}{\det Q},
    \qquad
    \prod_{k=1}^n\lambda_k=s^{-1}.
$$
Introducing a Lagrange multiplier \(\gamma\), we consider
$$
    \mathcal L =\sum_{k=1}^n\frac{w_k}{\lambda_k}- \gamma \left( \prod_{k=1}^n\lambda_k-s^{-1} \right).
$$
This gives the optimal allocation
$$
    \lambda_k^\ast  = w_k \left( \frac{s^{-1}}{\det W} \right)^{1/n}.
    \label{eq:lambda_CS_opt_s}
$$
The derivation is given in Appendix~\ref{app:compatible_case}. Substituting back, one obtains
\begin{equation}
    C_S^\ast[W]s^{-1/n}=n(\det W)^{1/n}.
    \label{eq:CS_opt_fixed_s}
\end{equation}
Thus, in the compatible case, the optimal precision cost is fixed by the weight volume $\det W$ and the sloppiness scale $s^{1/n}$. This result provides the compatible reference case. When $U=0$, the optimal Fisher geometry is completely determined by the weights. When $U\neq0$, the incompatibility contribution may favor a different Fisher-information allocation.

\subsection{Two parameters: incompatibility as an additive term}

For two parameters, take
$$
    Q=\operatorname{diag}(\lambda_1,\lambda_2),
    \qquad
    W=\operatorname{diag}(w_1,w_2),
    \qquad
    U=
    \begin{pmatrix}
        0 & u\\
        -u & 0
    \end{pmatrix}.
$$
The $C_T$ bound becomes
\begin{equation}
    C_T=\frac{w_1}{\lambda_1}+\frac{w_2}{\lambda_2}+2\frac{\sqrt{w_1w_2}|u|}{\lambda_1\lambda_2}.
    \label{eq:CT_2p_s}
\end{equation}
At fixed sloppiness, $\lambda_1\lambda_2=s^{-1}$, so the incompatibility term becomes the constant $ 2\sqrt{w_1w_2}|u|s$. Therefore, the incompatibility increases the value of the bound but does not change the optimal Fisher eigenvalue ratio. The optimized value is
\begin{equation}
    C_T^\ast s^{-1/2}=  2\sqrt{w_1w_2}\,\left(1+|u|s^{1/2} \right).
    \label{eq:CT_2p_opt_s}
\end{equation}
The derivation is given in Appendix~\ref{app:two_parameter_case}. For $W=\mathbf 1$, this becomes $C_T^\ast s^{-1/2}=2(1+|u|s^{1/2}).$ Thus, in two parameters, the incompatibility term is purely additive at fixed sloppiness. There is only one parameter plane, so there is no possibility of redistributing incompatibility among different planes.

The QGT feasibility condition gives $|u|s^{1/2}\leq1$, as shown in Appendix~\ref{app:QGT_feasibility}. However, this constraint does not change the optimal Fisher allocation: in two parameters there is only one parameter plane, so the incompatibility term is fixed once $s$ and $u$ are fixed.

\subsection{Three parameters: distribution-dependent incompatibility}

The three-parameter case is qualitatively different. Define the dual vector of the skew-symmetric matrix $U$ as
$$
    \mathbf u= (u_1,u_2,u_3)^T = (U_{23},U_{31},U_{12})^T.
$$
Here $u_k=U_{ij}$ is the incompatibility in the parameter plane $(i,j)$.

For $W=\mathbf 1$, the three-parameter expression can be written as
$$
    C_T=\sum_{k=1}^3\frac{1}{\lambda_k}+2 \sqrt{\frac{U_{12}^2}{\lambda_1^2\lambda_2^2}+\frac{U_{31}^2}{\lambda_3^2\lambda_1^2}+\frac{U_{23}^2}{\lambda_2^2\lambda_3^2}}.
$$
Using the exact three-parameter result derived in Eq.~\eqref{eq:CTCS_three_parameter_s_form}, the trade-off bound can be written as
$$
    C_T=C_S+ 2\sqrt c\,s^{2/3}\sqrt{G_3^{(F)}} ,
$$
so we obtain
$$
    C_T=s^{1/3}\sum_{k=1}^3\frac1{\tilde\lambda_k}+2\sqrt c\,s^{2/3}\sqrt{\sum_{k=1}^3\tilde u_k^2\tilde\lambda_k^2} .
$$

This expression is the key difference from the two-parameter case. Besides the total incompatibility $c$, the cost also depends on the distribution $\tilde u_k^2$ of the incompatibility among the three parameter planes. The above optimization describes the allocation favored by the $C_T$ cost. However, not every pair $(Q,U)$ is physically feasible. The corresponding QGT feasibility condition is $c\,s^{2/3}\sum_k\tilde u_k^2\tilde\lambda_k\leq1$, see Appendix~\ref{app:QGT_feasibility}.

To see this effect explicitly, consider the axisymmetric family
$$
    \tilde u_1^2=a, \qquad \tilde u_2^2=\tilde u_3^2=\frac{1-a}{2}, \qquad a\in\left[\frac13,1\right].
$$
The value $a=1/3$ corresponds to an isotropic incompatibility distribution, while $a=1$ corresponds to a fully concentrated distribution in the plane $(2,3)$. By symmetry, the optimal Fisher geometry satisfies $\tilde\lambda_2=\tilde\lambda_3$. Using $\tilde\lambda_1\tilde\lambda_2\tilde\lambda_3=1$, we write $\tilde\lambda_2=\tilde\lambda_3=1/\sqrt{\tilde\lambda_1}$. Then the bound becomes
$$
    C_T=s^{1/3}\left(\frac1{\tilde\lambda_1}+2\sqrt{\tilde\lambda_1}\right)+2\sqrt c\,s^{2/3}\sqrt{a\tilde\lambda_1^2+\frac{1-a}{\tilde\lambda_1}} .
$$



Without imposing the QGT feasibility constrain, the optimized value is non-increasing with the concentration parameter:
\begin{equation}
    C_T^\ast(a_2)\leq C_T^\ast(a_1),  \qquad a_2>a_1\geq\frac13.
    \label{eq:axisymmetric_monotonicity}
\end{equation}
The derivation is given in Appendix~\ref{app:axisymmetric_monotonicity}. Without imposing the QGT feasibility constraint, the optimized value is non-increasing with $a$ within the axisymmetric family. The QGT condition further restricts the physically allowed Fisher allocations. If the unconstrained optimum satisfies the QGT condition, the monotonicity result applies directly. Otherwise, the physical optimum is located on the QGT boundary.

This result should not be interpreted as saying that incompatibility is always better when it is more concentrated. Rather, concentration identifies a dominant incompatible parameter plane. The Fisher geometry can then be reshaped in a targeted way, increasing the Fisher area of that plane and reducing the corresponding contribution to $C_T$. Thus the relevant mechanism is Fisher-area allocation, not concentration by itself.

\section{A three-parameter qutrit model}

We now illustrate the geometric mechanism in a concrete three-parameter qutrit model to show how the distribution of incompatibility has a visible effect on the trade-off cost in a physical estimation problem.

We consider the \(SU(2)\) unitary encoding \cite{Candeloro2024}
$$ |\psi(B,\theta,\varphi;t)\rangle=e^{-iH(B,\theta,\varphi)t}|\psi_0\rangle,$$
where $H(B,\theta,\varphi)=BJ_{\vec n(\theta,\varphi)}$, with
$$\vec n(\theta,\varphi)=(\cos\theta\cos\varphi,\cos\theta\sin\varphi,\sin\theta).$$
The three parameters to be estimated are $(B,\theta,\varphi)$. For each randomly sampled initial qutrit state $|\psi_0\rangle$, we fix the model parameters $B=1$, $\theta=0.5$, $\varphi=0.8$, and $t=1$, and compute $Q$ and $U$, diagonalize $Q$, and transform $U$ into the corresponding Fisher eigenbasis. The choice $t=1$ is used as a representative working point, because the purpose of the example is not to optimize over the interrogation time, but to test the geometric decomposition across different probe states. We then evaluate $c$, $s$, $G$, $C_T-C_S$, $C_T$, and $C_H$. In this example, we write $G\equiv G_3^{(F)}$ for simplicity. 

\begin{figure*}[t]
    \centering
\includegraphics[width=0.9\textwidth]{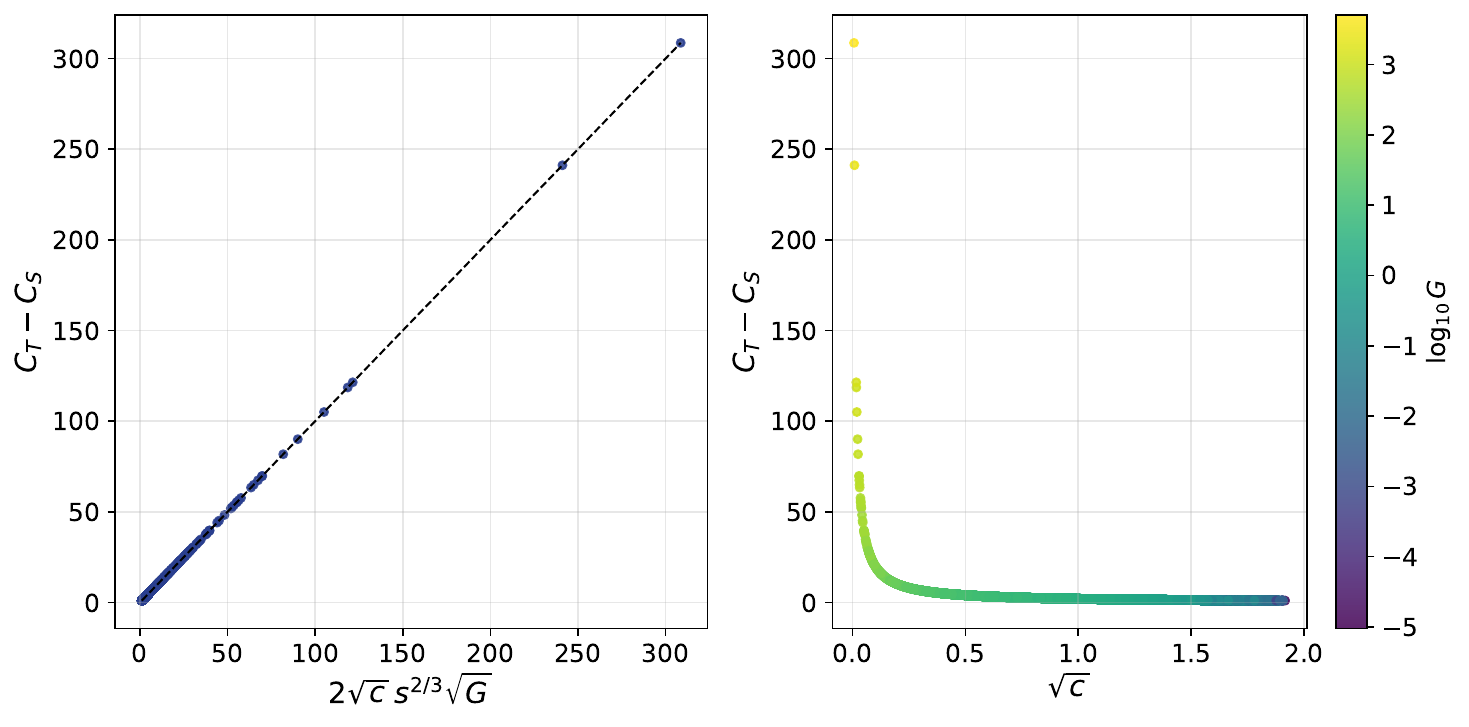}
    \caption{
    Numerical diagnosis of the incompatibility contribution in the three-parameter qutrit model. 
    The left panel verifies the exact three-parameter decomposition
    \(C_T-C_S=2\sqrt{c}\,s^{2/3}\sqrt{G}\).
    The right panel shows that the total incompatibility strength \(\sqrt c\) alone does not determine the incompatibility contribution; the color indicates \(\log_{10}G\).
    }
    \label{fig:qutrit_tradeoff_decomposition}
\end{figure*}

We first test the decomposition of the incompatibility contribution in Fig.~\ref{fig:qutrit_tradeoff_decomposition}. The left panel shows that the numerical data collapse onto the identity $C_T-C_S=2\sqrt c\,s^{2/3}\sqrt{G}$, confirming that the three-parameter formula derived above is realized in this physical qutrit model. The right panel shows that the total incompatibility strength $\sqrt c$ alone does not determine the penalty
$\|M\|_1=C_T-C_S$. In fact, the data contain regimes where $\sqrt c$ increases while $C_T-C_S$ decreases. This does not contradict the role of incompatibility; rather, it reflects the exact decomposition $C_T-C_S=2\sqrt c\,s^{2/3}\sqrt G$. A larger raw incompatibility can be less costly when it is accompanied by a smaller sloppiness scale or a smaller Fisher-area matching factor. We also find that larger $G$ generally leads to a larger incompatibility contribution and $G$ is not fixed by $\sqrt c$. Thus the amount of incompatibility and its placement relative to the Fisher geometry are distinct pieces of information. 

\begin{figure}[t]
    \centering
    \includegraphics[width=0.6\textwidth]{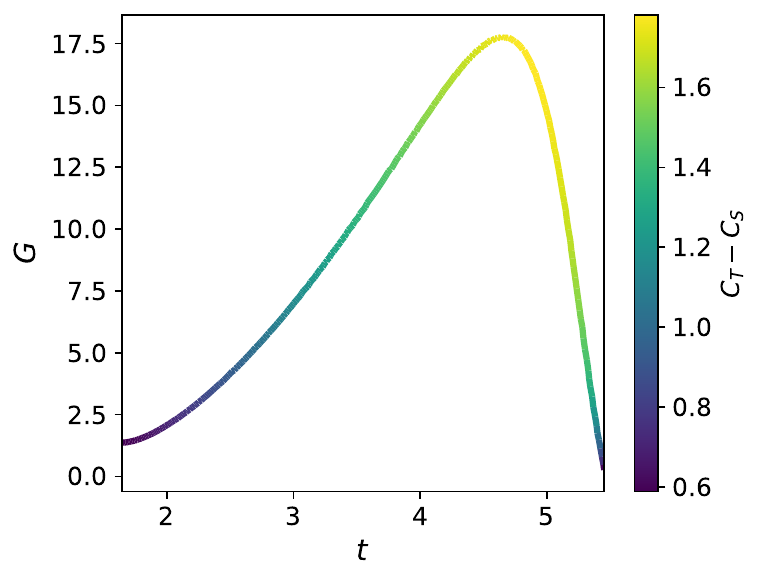}
    \caption{
Variation of $G$ along an approximately fixed-sloppiness slice in the $(t, \theta)$ control plane. The probe state and parametes $B$ and $\varphi$ are fixed, while $t$ and $\theta$ are varied and only points with $s\in[0.0198,0.0206]$ are retained. The color encodes the  value of the bound $C_T-C_S$. 
}
    \label{fig:qutrit_G_diagnostic}
\end{figure}

Since the factor $s^{2/3}$ can dominate the magnitude of $C_T-C_S$, the role of $G$ is most clearly seen when the sloppiness scale is controlled. We therefore examine whether $G$ can still vary along an approximately fixed-$s$ slice. This is important because fixing $s$ fixes the Fisher volume, but not the eigenvalue ratios of $Q$ nor the orientation of the incompatibility vector relative to the Fisher eigenbasis.
To illustrate that $G$ remains tunable after the sloppniess scale is fixed, we select an approximately constant-$s$ slice in the $(t,\theta)$ control plane. The probe state and the parameters $B$ and $\varphi$ are kept fixed, while the evolution time $t$ and the working angle $\theta$ are varied. We then retain only the points satisfying $s\in[0.0198,0.0206]$.
As shown in Fig.~\ref{fig:qutrit_G_diagnostic}, $G$ changes substantially within this narrow sloppiness window. This shows that, even after the Fisher volume is essentially fixed, the Fisher-area matching can still be modified by changing physical controls. In this sense, $G$ represents an additional optimization direction beyond the sloppiness scale: it captures how the incompatibility vector is placed relative to the Fisher eigenbasis. A smaller value of $G$ corresponds to a more favorable placement of incompatibility on parameter planes with larger Fisher area, thereby reducing the incompatibility contribution in $C_T-C_S$ when the other factors are comparable.

\section{Discussion and Conclusion}

\mga{In this work, we have discussed in detail the trade-off bound $C_T$ for multiparameter quantum estimation, highlighting the existence of two conceptually distinct ingredients: the total amount of incompatibility, measured by $c = \frac{1}{2}\|U\|_F^2$, and its distribution relative to the Fisher geometry, captured by the dimensionless matching factor $G_n^{(F)}$. While incompatibility has often been treated as a scalar resource, suggesting that more incompatibility simply means worse precision, our results show that this picture is incomplete. Two estimation problems with identical $c$ can exhibit very different $C_T$ bounds if their incompatibility is concentrated in different parameter planes. The crucial quantity is not $c$ alone, but rather how the Fisher-weighted incompatibility $M = Q^{-1}UQ^{-1}$ distributes its singular values.}

\mga{
Why does concentration sometimes help? The answer lies in the ability to reshape the Fisher geometry at fixed determinant. When incompatibility is isotropically spread across parameter planes, no single plane can be preferentially amplified. By contrast, when incompatibility is concentrated, say, in the $(2,3)$ plane, an optimal statistical model would have a reduced Fisher eigenvalue $\lambda_1$ (dual to that plane), thereby increasing the Fisher area $\lambda_2\lambda_3$ of the incompatible plane. This targeted geometric reshaping suppresses the cost term $\|M\|_1$. In other words, sloppiness becomes somehow a resource: a sufficiently anisotropic Fisher information can ''absorb'' the cost of incompatibility by allocating large Fisher area to the most problematic directions. This insight flips the conventional intuition that sloppiness is always detrimental.}

\mga{From a more applicative perspective, our analysis provides guidelines for optimization of statistical models over encoding mechanisms and probe states. In multiparameter quantum sensing, one should not simply minimize the total Uhlmann curvature or maximize the determinant of the QFIM. Instead, a more refined strategy is required: (i) identify which parameter planes carry the dominant incompatibility, (ii) engineer the probe state so that the corresponding Fisher eigenvalues are small (thereby increasing the Fisher area of those planes), and (iii) balance this against the compatible cost $\operatorname{Tr}(WQ^{-1})$. Our qutrit $SU(2)$ example explicitly demonstrates that probes leading to larger $c$ can outperform states with smaller one if the matching factor $G$ is sufficiently reduced. Thus, the relevant figure of merit for comparing probe states is not a scalar but a pair: $(C_S(\det Q)^{1/n},\, \jy{2\sqrt{c}\,(\det Q)^{-1/n}\sqrt{G}})$.}

\mga{Limitations of the present work should be acknowledged. First, our analysis assumes that the weight 
matrix $W$ is either the identity or simultaneously diagonal with $Q$. For non-diagonal $W$, the interplay between the cost metric and the Fisher geometry becomes richer and may introduce additional trade-offs. 
Second, our exact decomposition for three parameters relies on the special fact that a $3\times 3$ skew-symmetric matrix has a single nontrivial singular mode; for $n>3$, the trace norm $\|M\|_1$ receives contributions from multiple independent skew-symmetric modes, and the Frobenius bound~\eqref{eq:CTCS_bound_s_G} leaves a rank-dependent gap. Whether this gap can be saturated or further characterized remains open. Third, the qutrit example is restricted to pure states and a specific unitary encoding; exploring mixed states or different Hamiltonians would test the robustness of our conclusions. These limitations point to several future directions. One immediate extension is to understand the optimal distribution of incompatibility for $n>3$: should one concentrate all incompatibility into a single plane, or distribute it across a few planes with carefully chosen Fisher eigenvalues? This is an optimization problem over the set of feasible $(Q,U)$ pairs satisfying $Q + iU \ge 0$. Another direction concerns the role of measurements: the trade-off bound $C_T$ quantifies a fundamental limit, but it does not specify which measurements saturate it. \jy{Tight trade-off relations and saturating measurements for pure-state models have recently been studied from the viewpoint of attainable classical Fisher information~\cite{Wang2025, Wang2026}.} Understanding whether the optimal Fisher geometry derived here is actually attainable with physically realistic measurements remains an important open question. Our analysis is restricted to the second-order local geometry encoded in the QFIM and the incompatibility matrix. It would be interesting to investigate whether higher-order quantum geometric quantities can play a role in nonlinear or higher-order multiparameter estimation~\cite{hetenyi2023fluctuations}.}

\mga{In conclusion, we have shown that the precision cost of incompatibility in multiparameter quantum estimation is governed not by the total incompatibility strength but by its alignment with the Fisher eigenbasis. Concentration of incompatibility into a single parameter plane can reduce the optimized trade-off cost, provided the Fisher geometry is reshaped accordingly. The dimensionless matching factor $G_n^{(F)}$ provides a diagnostic that should become standard in probe-state design. More broadly, our work suggests that in multiparameter quantum metrology, sloppiness and incompatibility are not independent nuisances. Rather, they are geometric resources that can, when properly aligned, be made to work with rather than against each other.}

\appendix

\section{QGT feasibility conditions in low dimensions}
\label{app:QGT_feasibility}
We obtain the general QGT feasibility condition in Fisher-normalized form: $\sigma_{\max}\left(B\right)\leq1$.
We then show how this spectral condition reduces to simple determinant inequalities in two and three dimensions. However, for $n\geq 4 $, determinant positivity is no laonger sufficient.

\subsection{Two parameters}
For two parameters,
$$
    U=
    \begin{pmatrix}
        0 & u\\
        -u & 0
    \end{pmatrix}.
$$
In the eigenbasis of $Q$, let $Q=\operatorname{diag}(\lambda_1,\lambda_2)$.
Then
$$
    B=
    \begin{pmatrix}
        0 & \dfrac{u}{\sqrt{\lambda_1\lambda_2}}\\[1ex]
        -\dfrac{u}{\sqrt{\lambda_1\lambda_2}} & 0
    \end{pmatrix}.
$$
Thus the only nonzero singular value is
$$
    \beta= \frac{|u|}{\sqrt{\lambda_1\lambda_2}}=\frac{|u|}{\sqrt{\det Q}}.
$$
The condition $\beta\leq1$ gives
\begin{equation}
    \det Q\geq u^2 .
    \label{eq:app_QGT_2p_condition}
\end{equation}
In terms of the sloppiness $s=1/\det Q$, this can also be written as
$$ |u|s^{1/2}\leq1.$$

\subsection{Three parameters}

In the eigenbasis of $Q$, $Q=\operatorname{diag}(\lambda_1,\lambda_2,\lambda_3)$,
the entries of $B$ are $B_{ij}=\frac{U_{ij}}{\sqrt{\lambda_i\lambda_j}}$.
A $3\times3$ real skew-symmetric matrix has only one nonzero singular-value pair. Its squared singular value is
$$\beta^2=\frac{\lambda_1U_{23}^2+ \lambda_2U_{31}^2 + \lambda_3U_{12}^2  }{\det Q}.$$
With $\mathbf u=(U_{23},U_{31},U_{12})^T$, this can be written as
$$\beta^2=\frac{\mathbf u^TQ\mathbf u}{\det Q}.$$
Therefore the QGT feasibility condition $\beta\leq1$ gives
\begin{equation}
    \det Q\geq \mathbf u^TQ\mathbf u .
    \label{eq:app_QGT_3p_condition}
\end{equation}

Using the sloppiness $s=1/\det Q$, the dimensionless Fisher eigenvalues $\tilde\lambda_k=s^{1/3}\lambda_k$, and $c=|\mathbf u|^2$, this condition becomes
   $$ c\,s^{2/3} \sum_{k=1}^3 \tilde u_k^2\tilde\lambda_k \leq1,$$
where $\tilde u_k^2=u_k^2/|\mathbf u|^2$. 

For the axisymmetric family $\tilde u_1^2=a$, $\tilde u_2^2=\tilde u_3^2=\frac{1-a}{2}$,
and with $\tilde\lambda_2=\tilde\lambda_3 = \frac{1}{\sqrt{\tilde\lambda_1}}$,
the feasibility condition reduces to
\begin{equation}
    c\,s^{2/3}
    \left( a\tilde\lambda_1+\frac{1-a}{\sqrt{\tilde\lambda_1}}\right)\leq1.
    \label{eq:app_QGT_axisymmetric_condition}
\end{equation}
This condition restricts the Fisher-information redistribution allowed by the $C_T$ optimization. If the unconstrained optimum violates Eq.~\eqref{eq:app_QGT_axisymmetric_condition}, the physical optimum must lie on the QGT boundary.

\subsection{Why determinant positivity is insufficient for $n \geq 4$}

The determinant of the QGT can also be expressed in terms of the singular values of $B$, and we have
$$\det(Q+iU)=\det Q\,\det(I+iB).$$
Using the canonical block form of a real skew-symmetric matrix, each nonzero singular-value block contributes a factor
$$\det(I+iB_\alpha)= (1+\beta_\alpha)(1-\beta_\alpha)=1-\beta_\alpha^2.$$
Therefore,
\begin{equation}
    \det(Q+iU)= \det Q\prod_{\alpha=1}^{\lfloor n/2\rfloor} (1-\beta_\alpha^2).
    \label{eq:app_det_QGT_general}
\end{equation}

For $n=2$ and $n=3$, there is only one nonzero singular-value mode. In these cases, the condition $\det(Q+iU)\geq0$ is equivalent to $\beta_1\leq1$, and therefore reproduces the feasibility conditions above.

For $n\geq4$, however, there can be more than one nonzero mode. Then $\prod_\alpha(1-\beta_\alpha^2)\geq0$ does not imply $\beta_\alpha\leq1$ for every $\alpha$.
For example, if two singular values satisfy $\beta_1>1$ and $\beta_2>1$, then both factors $1-\beta_1^2$ and $1-\beta_2^2$ are negative, so their product is positive, even though $I+iB$ has negative eigenvalues. 

\section{Fixed-sloppiness optimizations}
\label{app:fixed_sloppiness}

\subsection{Compatible case}
\label{app:compatible_case}

We derive Eq.~\eqref{eq:CS_opt_fixed_s}. In the basis where $Q$ and $W$ are diagonal,
    $$C_S[W]=\sum_{k=1}^n\frac{w_k}{\lambda_k},
    \qquad
    \prod_{k=1}^n\lambda_k=s^{-1}.$$
Consider the Lagrangian
    $$\mathcal L= \sum_{k=1}^n\frac{w_k}{\lambda_k}-\gamma \left(\prod_{k=1}^n\lambda_k-s^{-1}\right).$$
At the stationary point, using $\prod_j\lambda_j=s^{-1}$, we have
$$\frac{\partial\mathcal L}{\partial\lambda_k} =-\frac{w_k}{\lambda_k^2}- \gamma\frac{s^{-1}}{\lambda_k} = 0.$$
Thus $\lambda_k\propto w_k$. Writing $\lambda_k=\kappa w_k$ and imposing the constraint gives
$$\kappa^n\det W=s^{-1},\qquad\kappa=\left(\frac{s^{-1}}{\det W}\right)^{1/n}.$$
Therefore
$$\lambda_k^\ast =  w_k \left( \frac{s^{-1}}{\det W} \right)^{1/n}.$$
Substituting into $C_S[W]$ gives
$$C_S^\ast[W]=\sum_{k=1}^n\frac{w_k}{w_k(s^{-1}/\det W)^{1/n}}=n(\det W)^{1/n}s^{1/n}.$$

\subsection{Two-parameter case}
\label{app:two_parameter_case}

For two parameters,
$$
    C_T=  \frac{w_1}{\lambda_1}+  \frac{w_2}{\lambda_2}  +  2  \frac{\sqrt{w_1w_2}|u|}{\lambda_1\lambda_2}.
$$
At fixed sloppiness, $\lambda_1\lambda_2=s^{-1}$. Hence the incompatibility term becomes $2\sqrt{w_1w_2}|u|s$, which is independent of the ratio $\lambda_2/\lambda_1$. The remaining minimization is
$$
    \min_{\lambda_1\lambda_2=s^{-1}} \left(\frac{w_1}{\lambda_1}+\frac{w_2}{\lambda_2}\right).
$$
Using $\lambda_2=s^{-1}/\lambda_1$, this becomes
$$
    \frac{w_1}{\lambda_1}+w_2s\lambda_1.
$$
The stationary condition gives
$$-\frac{w_1}{\lambda_1^2}+w_2s=0,$$
and therefore
$$\lambda_1^\ast=\sqrt{\frac{w_1}{w_2}}s^{-1/2},\qquad\lambda_2^\ast=\sqrt{\frac{w_2}{w_1}}s^{-1/2}.$$
The optimized value is
$$C_T^\ast=2\sqrt{w_1w_2}s^{1/2}+2\sqrt{w_1w_2}|u|s=2\sqrt{w_1w_2}s^{1/2}\left(1+|u|s^{1/2}\right).$$

\subsection{Axisymmetric three-parameter monotonicity}
\label{app:axisymmetric_monotonicity}

We prove Eq.~\eqref{eq:axisymmetric_monotonicity}. For the axisymmetric family, $\tilde u_1^2=a$ and $\tilde u_2^2=\tilde u_3^2=(1-a)/2$. With $\tilde\lambda_2=\tilde\lambda_3=1/\sqrt{\tilde\lambda_1}$, the cost becomes
$$ C_a(\tilde\lambda_1):=C_T=s^{1/3}\left(\frac{1}{\tilde\lambda_1}+2\sqrt{\tilde\lambda_1}\right)+2\sqrt c\,s^{2/3}\sqrt{a\tilde\lambda_1^2+\frac{1-a}{\tilde\lambda_1}}.$$
For fixed $\tilde\lambda_1$, the only $a$-dependent part is the expression inside the square root, and
$$\frac{\partial}{\partial a}\left(a\tilde\lambda_1^2+\frac{1-a}{\tilde\lambda_1}\right)=\tilde\lambda_1^2-\frac{1}{\tilde\lambda_1}.$$
Thus, for $0<\tilde\lambda_1<1$, increasing $a$ decreases the incompatibility contribution. Moreover,
$$\left.\frac{\partial C_a}{\partial\tilde\lambda_1}\right|_{\tilde\lambda_1=1}=\sqrt c\,s^{2/3}(3a-1).$$
Therefore, for $a>1/3$, moving left from $\tilde\lambda_1=1$ lowers the cost, and the unconstrained optimum satisfies $\tilde\lambda_1^\ast<1$.

Let $a_2>a_1\geq1/3$, and let $\tilde\lambda_{1,a_1}^\ast$ be the optimizer for $a_1$. Since $\tilde\lambda_{1,a_1}^\ast\leq1$, we have
$$C_{a_2}(\tilde\lambda_{1,a_1}^\ast)\leq C_{a_1}(\tilde\lambda_{1,a_1}^\ast)=C_{a_1}^\ast.$$
Since $C_{a_2}^\ast$ is the minimum of $C_{a_2}$,
$$C_{a_2}^\ast\leq C_{a_2}(\tilde\lambda_{1,a_1}^\ast)\leq C_{a_1}^\ast.$$
Thus
$$C_T^\ast(a_2)\leq C_T^\ast(a_1),\qquad a_2>a_1\geq\frac13.$$
This proves the monotonicity within the axisymmetric family.

\bibliographystyle{unsrt}
\bibliography{sample.bib}

\end{document}